\documentclass[12pt,letterpaper]{article}
\usepackage{jheppub}
\usepackage{verbatim}

\usepackage{graphicx}
\usepackage{subfigure}
\input{epsf}
\usepackage{epsfig}
\usepackage{epstopdf}
\usepackage{amsthm}
\usepackage{mathrsfs}
\usepackage{MnSymbol}

\newcommand{\be}{\begin{equation}}
\newcommand{\ee}{\end{equation}}
\def\({\left(} \def\){\right)}
\def\[{\left[} \def\]{\right]}

\title{}
\def\sgn{\text{sgn}}

\def\gsim{\, \rlap{$>$}{\lower 1.1ex\hbox{$\sim$}}\,}
\def\lsim{\, \rlap{$<$}{\lower 1.1ex\hbox{$\sim$}}\,}

\newcommand{\bea}{\begin{eqnarray}}
\newcommand{\eea}{\end{eqnarray}}

\title{The Spectrum in the Sachdev-Ye-Kitaev Model}

\author{Joseph Polchinski}
\author{and Vladimir Rosenhaus}
\affiliation{Kavli Institute for Theoretical Physics, \\
 University of California, Santa Barbara, CA 93106}

\emailAdd{joep@kitp.ucsb.edu, vladr@kitp.ucsb.edu}

\abstract{ The SYK model consists of $N\gg 1$ fermions in $0+1$ dimensions with a random, all-to-all quartic interaction. Recently, Kitaev has found that the SYK model is maximally chaotic and has proposed it as a model of holography. We solve the Schwinger-Dyson equation and compute the spectrum of two-particle states in SYK, finding both a continuous and discrete tower. The  four-point function is expressed as a sum over the spectrum. The sum over the discrete tower is evaluated. 
}

\begin{document}
\maketitle
\section{Introduction}
The Sachdev-Ye-Kitaev model  (SYK) \cite{Kitaev, SY} is a $0+1$ dimensional model of $N\gg1$ fermions with an  all-to-all random quartic interaction. SYK has three notable features:
\begin{description}
\item[Solvable at strong coupling.] At large $N$ one can sum all Feynman diagrams, and thereby compute correlation functions at strong coupling. 
\item [Maximally chaotic.] Chaos is quantified by the Lyapunov exponent, which is defined by an out-of-time-order four-point function \cite{LO, KitaevNov}. The Lyapunov exponent of a black hole in Einstein gravity is $2\pi/\beta$ \cite{Shenker:2013pqa,  Shenker:2014cwa, KitaevNov},  where $\beta$ is the inverse temperature. This is the maximal allowed Lyapunov exponent \cite{Maldacena:2015waa}, and  SYK saturates the bound \cite{Kitaev}. 
\item[ Emergent conformal symmetry.] In the context of the two-point function, there is emergent conformal symmetry at low energies \cite{PG, Sachdev:2010um, Sachdev:2010uj, Kitaev}.
\end{description}

Due the scarcity of nontrivial systems which can be solved at strong coupling, the first item is already enough to make the model worthy of study. The combination of the first and the second items is remarkable and surprising. In the context of classical systems, solvability usually means integrability, which is mutually exclusive from chaos. For a quantum system, there is no such restriction, as  SYK  demonstrates. The third item implies that the model has some kind of holographic dual. The second item strongly suggests this dual is Einstein gravity in some form. The combination of all three items would appear to potentially place the model in the unique class of constituting a solvable model of holography. 

SYK is a variant of the Sachdev-Ye model (SY) \cite{SY} that was introduced by Kitaev in a series of seminars \cite{Kitaev}. Kitaev made significant advances in understanding the model, connected the holographic study of chaos of Shenker and Stanford \cite{Shenker:2013pqa, Shenker:2013yza, Shenker:2014cwa} to Lyapunov exponents \cite{LO}, and proposed SYK as a model of holography.

 The main goal of this paper is to study the four-point function. This is also being considered in \cite{MStoappear, Ktoappear}.
In Section \ref{sec:2pt} we review the model, its two-point function, and the emergent conformal symmetry. 
In Section \ref{sec:eigen} we first review the setup of the four-point function introduced in \cite{Kitaev}. We then proceed to solve the Schwinger-Dyson equation to compute the spectrum of two-particle states. We find both a discrete tower and a continuous tower. In Section \ref{sec:4pt} the four-point function is expressed as a sum over the spectrum. The discrete part of the sum is explicitly evaluated. Some comments are made on the breaking of conformal invariance.

\section{Two-point function} \label{sec:2pt}
\begin{figure} 
\centering
\subfigure[]{
\includegraphics[width=2.5in]{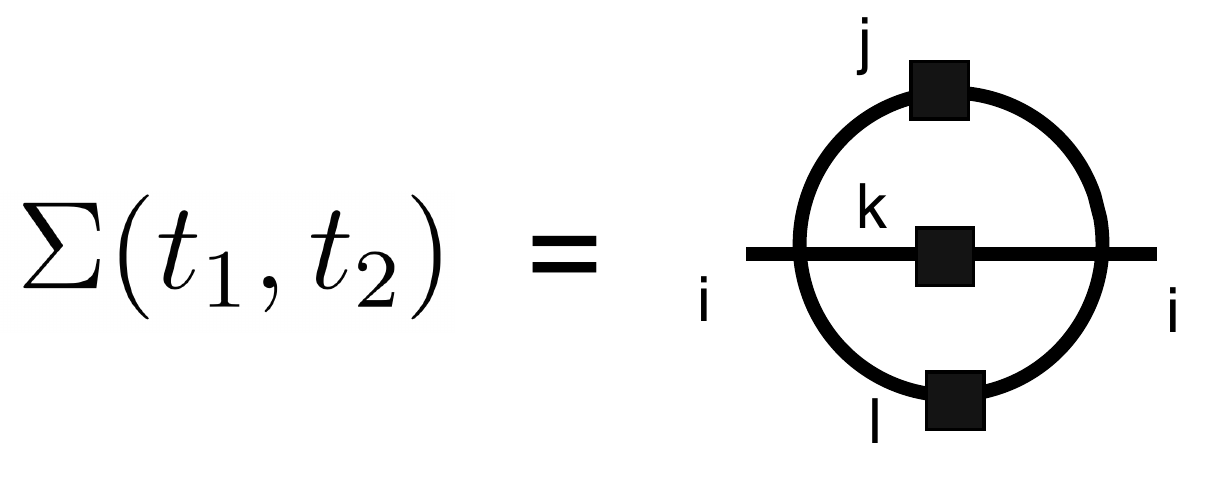}
}
\subfigure[]{
\includegraphics[width=6in]{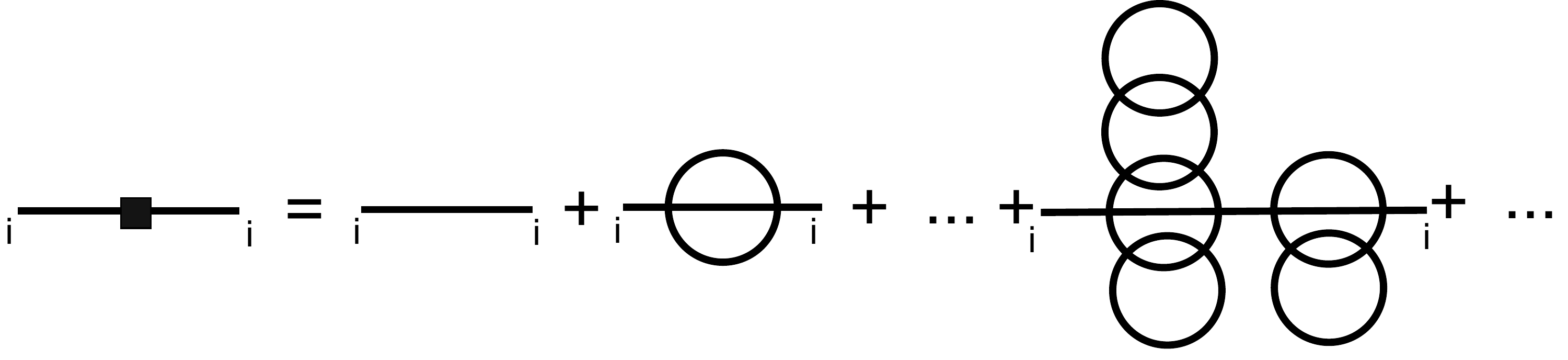}
}
\subfigure{
\includegraphics[width=5in]{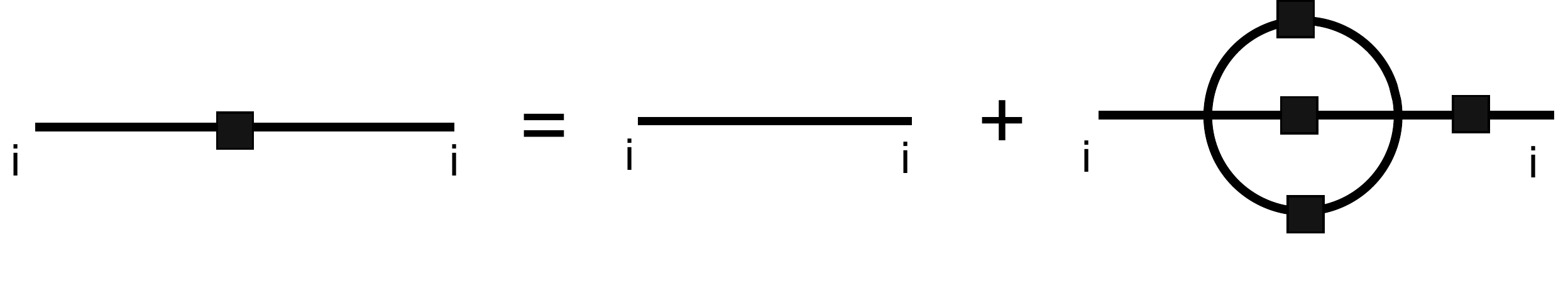}
}
\caption{The line with a box is the full two-point function, while the solid line is the free two-point function. (a) The self-energy $\Sigma(t_1, t_2)$ in terms of the two-point function $G(t_1, t_2)$. (b) Some of the Feynman diagrams making up the two-point function. (c) The Schwinger-Dyson equation for the two-point function. Iterating generates the sum in (b). } \label{fig:SD2pt}
\end{figure}

The SYK model is given by the Hamiltonian \cite{Kitaev}, 
\be \label{HSYK}
H =\frac{1}{4!} \sum_{i,j,k,l = 1}^N J_{i j k l}\ \chi_i \chi_j \chi_k \chi_l~,
\ee
where $\chi_j$ are Majorana fermions $\{ \chi_i, \chi_j\} =  \delta_{i j}$, and the model has quenched disorder with the couplings $J_{i j k l}$ drawn from the distribution, 
\be
P(J_{i j k l}) \sim \exp\(- N^3 J_{i j k l}^2/ 12 J^2\)~,
\ee
leading to a disorder average of,
\be
\overline{J_{i j k l}^2} = \frac{3! J^2}{N^3}~, \ \ \ \ \overline{J_{i j k l}} = 0~.
\ee
The expressions for the correlation functions that will follow will always be the result after the disorder average has been performed. 
The Lagrangian trivially follows from the Hamiltonian and is,
\be
L = -\frac{1}{2}\ \chi_j \frac{d}{d t}\chi_j - H~.
\ee
The couplings $J_{i j k l}$ have dimension $1$, while the  fermions $\chi_i$ have dimension $0$. 
The free two-point function for the fermions is,
\be \label{G0}
G_0(t)  \delta_{i j} \equiv -\langle T \chi_i (t) \chi_j (0)\rangle = - \frac{1}{2} \sgn(t) \delta_{i j}~.
\ee
As a result of the disorder average, the anticommutation of the fermions, and large $N$, the Feynman diagrams for the  full  (zero temperature) two-point function take a remarkably simple form. The self energy $\Sigma(t_1, t_2)$ (1PI) is expressed in terms of the two-point function $G(t_1,t_2)$ (see  Fig.~\ref{fig:SD2pt}a) 
\be \label{eq:Sig}
\Sigma(t_1, t_2) = J^2 G(t_1, t_2)^3~.
\ee
Expressing the two-point function as a sum of the 1PI diagrams, 
\be \label{eq:G}
G ( i \omega)^{-1} = i \omega - \Sigma(i \omega)~.
\ee
The equations (\ref{eq:Sig}) and (\ref{eq:G}) fully determine the two-point function. Their solution is only known in the limit of low energies. In this limit, one may drop the $i \omega $ in (\ref{eq:G}), leading the Fourier transform of (\ref{eq:G}) to become
\be \label{GSig}
\int d t\ G(t_1, t) \Sigma(t, t_2) =  -\delta(t_1 - t_2)~.
\ee
Combining (\ref{GSig}) with (\ref{eq:Sig}) gives an integral equation for $G(t_1, t_2)$,
\be \label{eq:Int2pt}
J^2 \int d t\ G(t_1, t) G(t, t_2)^3 =  -\delta(t_1 - t_2)~,
\ee
which one can check is solved by \cite{SY},
\be \label{eq:2pt}
G(t) =- \(\frac{1}{4 \pi J^2}\)^{1/4} \frac{1}{\sqrt{|t|}} \sgn(t)~.
\ee 
The solution (\ref{eq:2pt}) for the Euclidean two-point function is valid at low energies, or equivalently, at strong coupling: the time separation $t$ should satisfy $J |t| \gg 1$. 
On the basis of the two-point function, it appears that the theory flows to an IR conformal fixed point, with the fermions acquiring an anomalous dimension $\Delta = 1/4$. The  above equations (\ref{eq:Sig}, \ref{eq:G}) determining the two-point function can either be found from the Feynman diagrams, as has been done here following Ref.~\cite{Kitaev}, or equivalently by performing the disorder average via the replica trick and evaluating the saddle point of the action \cite{SY,  Sachdev:2015efa}. 

An equivalent way to find the two-point function is from the Schwinger-Dyson equation in the form (see Fig.~\ref{fig:SD2pt}c),
\be \label{SD2pt}
G(t) = G_0(t) + J^2 \int d t_1 d t_2\ G_0(t_1) G(t_1,t_2)^3 G(t_2, t)~.
\ee
In the IR, one may drop the left-hand side, and find the solution (\ref{eq:2pt}). 
The late time decay of $G(t)$, as compared to the constant behavior of $G_0(t)$, demonstrates that dropping that left-hand side in (\ref{SD2pt}) was self-consistent.

To go to finite temperature one uses the conformal invariance of the Schwinger-Dyson equation (\ref{eq:Int2pt}) \cite{PG, Kitaev}. Suppose $G(\sigma_1,\sigma_2)$ solves (\ref{eq:Int2pt}), 
\be \label{eq:Int2pt2}
J^2 \int d \sigma \ G(\sigma_1, \sigma) G(\sigma, \sigma_2)^3 =  -\delta(\sigma_1 - \sigma_2)~.
\ee
Consider an arbitrary time reparameterization, $\sigma = f(t)$. One can check that (\ref{eq:Int2pt2}) transforms into (\ref{eq:Int2pt}) provided one lets 
\be \label{Gconf}
G(t_1, t_2) = |\partial_1 f(t_1) \partial_2 f (t_2)|^{1/4} G(\sigma_1, \sigma_2)~.
\ee
Choosing $f(t) = e^{2\pi i t/\beta}$ maps the line into a circle, transforming the zero-temperature two-point function into a finite-temperature two-point function \cite{PG}, 
\be
G_{\beta}(t) = - \frac{\pi^{1/4}}{\sqrt{2  \beta J}} \frac{1}{\sqrt{\sin(\pi t/\beta)}}~,
\ee
where the temperature is $\beta^{-1}$. Analytically continuing to real time $t = - i t_r$ turns $\sin(\pi t_r/\beta)$ into $\sinh(\pi t_r/\beta)$, giving an exponential late time decay of the thermal two-point function, as is expected for a strongly coupled CFT.

\subsubsection*{Sachdev-Ye}

The SYK model is closely related to the Sachdev-Ye model (SY), which we now review. SY involves  $N\gg 1 $ spins with Gaussian-random, infinite-range exchange interactions \cite{SY},
\be \label{HSY}
H = \frac{1}{\sqrt{M}}\sum_{j, k=1}^N J_{j k}\ \vec{S_j} \cdot \vec{S_k}~,
\ee
where the $J_{i j}$ are drawn from the distribution,
\be
P(J_{i j}) \sim \exp(- J_{ij}^2/ 2 J^2)~,
\ee
and the spins are in some representation of $SU(M)$. The choice of $SU(2)$ was studied by Bray and Moore \cite{BM}, and it was numerically found to have  spin-glass order at zero temperature.  Sachdev and Ye \cite{SY} considered (\ref{HSY}) in an arbitrary representation of $SU(M)$,  obtaining analytic control over (\ref{HSY}) in the limit $M\gg 1$. The correlators in SY are obtained by representing the spins in terms of fermions \cite{SY},
\be \label{eqS}
S_{\mu}^{\nu} = c^{\dagger}_{\mu} c^{\nu}, \ \ \ \ \ \ \ \sum_{\mu} c^{\dagger}_{\mu} c^{\mu} = n_b~,
\ee
where $n_b$ denotes the number of columns in the Young tableaux characterizing the representation of $SU(M)$, and (\ref{eqS}) holds at each site. Under the mapping (\ref{eqS}) the Hamiltonian (\ref{HSY}) is transformed into,
\be \label{HSY2} 
H=\frac{1}{\sqrt{M}} \sum_{i, j=1}^{N}\sum_{\mu, \nu =1}^{M} J_{ i j}\  c^{\dagger}_{i \mu} c^{\dagger}_{j \nu}c_i^{\nu} c_j^{\mu}~,
\ee
which, like the SYK Hamiltonian (\ref{HSYK}), is quartic in the fermions. Depending on the representation of $SU(M)$, the ground state may or may not be a spin glass. One choice of representation which was shown in \cite{GPS} to avoid a spin glass phase is one with a Young tableaux that has $n_b= \mathcal{O}(1)$ columns and $\mathcal{O}(M)$ rows, where $M\gg 1$ \cite{SY}. 
In order to a have a system that can serve as a model of holography, it is important that there not be a spin-glass phase \cite{KitaevNov}.~\footnote{A maximal Lyapunov exponent \cite{Maldacena:2015waa} could potentially occur in the highly quantum regime, at low temperatures. It is therefore important that the system not freeze into a spin glass as the temperature is lowered.} For SYK, a spin glass phase is manifestly avoided, as the fermions can not condense at a site (unlike the case of SY where the the fermions have an additional gauge index $\mu$) \cite{Sachdev:2015efa}. SYK is simpler than SY, in that it only requires a single scaling limit $N\rightarrow \infty$, while SY also requires $M\rightarrow \infty$. On the other hand, it may be useful to study SY as well, as it has a 2-index coupling, which may fit better with a bulk string theory than the 4-index coupling  $J_{ i j k l}$ in SYK.

\section{Spectrum} \label{sec:eigen}

\begin{figure} 
\centering
\subfigure[]{
\includegraphics[width=3in]{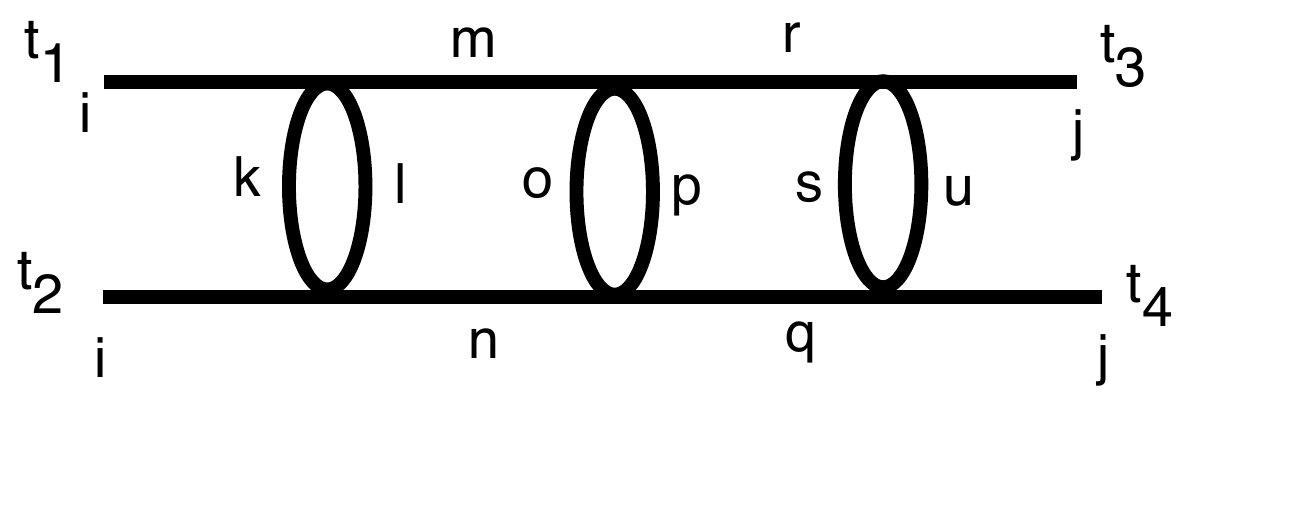}
}
\subfigure[]{
\includegraphics[width=6.4in]{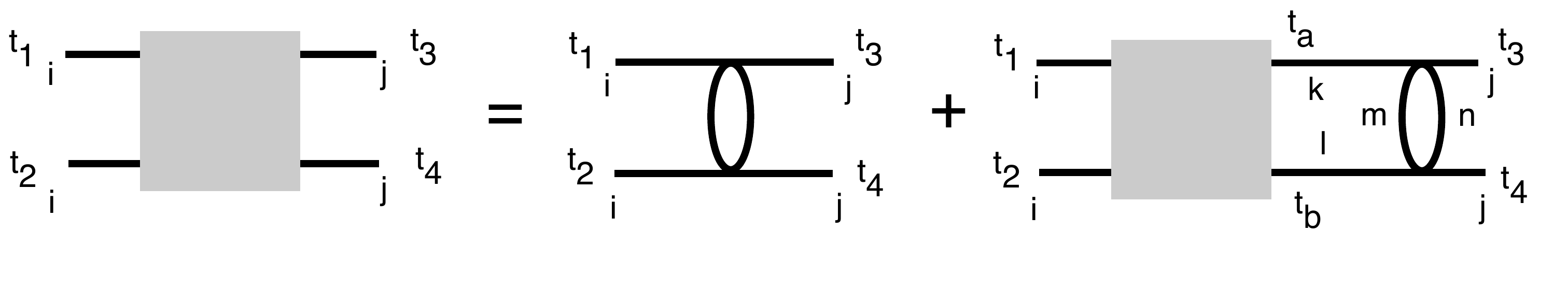}
}
\caption{ (a) The four-point function is given by a sum of  ladder diagrams, such as the one shown. (b) These ladder diagrams are generated by iterating the Schwinger-Dyson equation (note: the propagators are really the dressed propagators; we have suppressed the box on the line that it is meant to indicate this).} \label{fig:SD4pt}
\end{figure}

In this section we turn to the study of the four-point function,
\be  \label{eq:4chi}
\langle  \chi_i (t_1)  \chi_i (t_2) \chi_j (t_3) \chi_j (t_4)  \rangle~.
\ee 
The leading order connected piece scales as $1/N$. As with the two-point  function, the large $N$ structure of the four-point function is remarkably simple. At leading order, it is given entirely by the ladder diagrams shown in Fig.~\ref{fig:SD4pt} \cite{Kitaev}. 

The 1PI four-point function  satisfies the Schwinger-Dyson equation (Fig.~\ref{fig:SD4pt}b), 
\be \label{eq:Gamma}
\Gamma(t_1, t_2, t_3, t_4) =  \Gamma_0(t_1,t_2,t_3,t_4)  + \int d t_a  d t_b\ \Gamma(t_1, t_2, t_a, t_b) K(t_a, t_b, t_3, t_4)~,
\ee 
where
\bea 
K(t_a, t_b, t_3, t_4) &=& - 3 J^2 G(t_a, t_3)G(t_b, t_4)G(t_3, t_4)^2 \,,\nonumber\\ \label{KGamma0}
\Gamma_0(t_1,t_2,t_3,t_4) &=& 3 J^2 \delta(t_{13}) \delta(t_{24}) G(t_1,t_2)^2 \,,
\eea
and $G(t_1,t_2)$ is the two-point function (\ref{eq:2pt}), and we sometimes use the notation $t_{ij} = t_i - t_j$. Finding the four-point function amounts to solving the integral equation (\ref{eq:Gamma}). Regarding the kernel $K(t_a,t_b, t_3, t_4)$ as a matrix $\langle t_a\ t_b| K| t_3\ t_4\rangle$, a straightforward way to solve (\ref{eq:Gamma}) is by diagonalizing the kernel. The goal of this section will be to compute the eigenvectors $v(t_a, t_b)$ of the kernel. The four-point function will then  follow, and will be discussed in Sec.~\ref{sec:4pt}.

Some of the eigenvectors can be found by assuming a form that is a power of the time separation $t_{a b}$. In Sec.~\ref{sec:eigenvalues1} we review Kitaev's calculation of the eigenvalues of the kernel for this set. Surprisingly, there is an $SL(2,\mathcal{R})$ symmetry in the $t_a, t_b$ space. This was recognized by Kitaev, and is a hint of the holographic nature of SYK: the bulk $AdS_2$ is a hyperboloid in embedding coordinates, having the symmetry $SO(2,1) \sim SL(2,\mathcal{R})$. In Sec.~\ref{sec:sl2} we exploit this insight and use the $SL(2,\mathcal{R})$ symmetry to generate all the eigenvectors. Subtleties associated with boundary terms are discussed in Appendix~\ref{appendixB}. In Sec.~\ref{sec:direct} we directly verify that these are eigenvectors of the kernel. In Sec.~\ref{complete} we find the basis of eigenvectors that span the $t_a, t_b$ space. 

\subsection{Eigenvalues} \label{sec:eigenvalues1}
To find the spectrum of the theory, we must solve for the eigenvalues $g(\alpha)$ and eigenvectors $v_{\alpha }(t_a,t_b)$ of the kernel, 
\be  \label{eq:eig}
\int d t_a d t_b\ v_{\alpha}(t_a,t_b )\ K(t_a, t_b, t_3, t_4) = g(\alpha)\  v_{\alpha}(t_3, t_4)~.
\ee
Schematically, we can write (\ref{eq:eig}) as,
\be
K  v_{\alpha} = g(\alpha)  v_{\alpha} \,\label{eigen2} \,.
\ee
One set of eigenvectors that satisfy (\ref{eq:eig}) are \cite{Kitaev},
\be \label{eq:v}
v_{\alpha} (t_a, t_b) = \frac{1}{|t_a - t_b|^{2 \alpha}}\ \sgn(t_a - t_b)~.
\ee
The corresponding eigenvalues $g(\alpha)$ are found by plugging $v_{\alpha}$ into the equation (\ref{eq:eig}). 
The integral on the left-hand side of (\ref{eq:eig}) is,
\be \label{eigint1}
\int d t_a \int d t_b \frac{ \sgn(t_a - t_b)}{|t_a - t_b|^{2 \alpha}}\ \frac{\sgn(t_a - t_3)}{|t_a - t_3|^{1/2}}\ \frac{\sgn(t_b - t_4)}{|t_b - t_4|^{1/2}}~.
\ee
There are $8$ regions of integration, arising from each of the $\sgn$'s being positive or negative, which must be done separately; the computation is performed in Appendix~\ref{appendixA}. 
The result is \cite{Kitaev},
\be \label{eq:g}
g(\alpha) = - \frac{3}{2} \frac{1}{(1- 2 \alpha) \tan (\pi \alpha)}~.
\ee
In fact, the integral (\ref{eq:eig}) is divergent for all $\alpha$, and the result (\ref{eq:g}) implicitly involved analytic continuation.~\footnote{For instance, one of the regions of integration, region $7$ in the notation of Appendix~\ref{appendixA}, which is for $t_a<t_b, t_a<t_3, t_b>t_4$ , gives a result which is zero. This is for an integral of a manifestly positive quantity. A result of zero arises because the contributions to this integral, (\ref{eq:7a}) and (\ref{eq:7b}), precisely cancel.} We will have a better understanding of this divergence once we find the complete set of eigenvectors.

\subsection{$SL(2,\mathcal{R})$ and all eigenvectors} \label{sec:sl2}
We now use the eigenvectors (\ref{eq:v}) and the $SL(2,\mathcal{R})$ algebra to generate all the eigenvectors. 

Consider the $SL(2,\mathcal{R})$ algebra with the standard generators $L_p$ ,
\bea
L_p = t_1^p \partial_{t_1} + t_2^p \partial_{t_2}\,,\quad p = 0, 1, 2 \,,
\nonumber\\
{}[L_p, L_q] = (q-p) L_{p+q-1} \,.
\eea
One can perform a similarity transform to find new generators which also satisfy the same $SL(2,\mathcal{R})$ algebra. It will be useful to define $\tilde L_p = |t_{12}|^{-3/2} L_p |t_{12}|^{3/2}$, so that
\be
\tilde L_0 = L_0\,,\quad \tilde L_1 = L_1 + \frac32\,,\quad \tilde L_2 = L_2 + \frac32(t_1 + t_2) \,.
\ee
The advantage of the $\tilde L_p$ is that, at least naively, one finds they commute with the kernel,
\be
\tilde L_p K = K \tilde L_p \label{lkkl}~,
\ee
in the notation of~(\ref{eigen2}). So, the $\tilde L_p$ take solutions of~(\ref{eigen2}) to new solutions with the same eigenvalue. In fact, this statement is subtle and requires a careful treatment of boundary terms, and we elaborate more on it in Appendix~\ref{appendixB}. 

We can generate new solutions with $\tilde{L} _2$,
\be
\partial_\lambda v_{\alpha\lambda}(t_1,t_2) = (t_1 t_2)^{-3/2} \tilde{L}_2 (t_1 t_2)^{3/2} v_{\alpha\lambda}(t_1,t_2) \,.
\ee
Integrating this with the initial condition~(\ref{eq:v}) gives \cite{Kitaev},
\be
v_{\alpha\lambda }(t_1,t_2) = |1 - \lambda t_1|^{2\alpha-3/2} |1 - \lambda t_2|^{2\alpha-3/2} \frac{\sgn(t_{12})}{|t_{12}|^{2\alpha}} \,.
\ee
This would be an acceptable set, but it is better to take a set of definite frequency (which are distinguished by their subscript),
\be
v_{\alpha\omega}(t_1,t_2) = \int_{-\infty}^\infty d\tau e^{-i\omega \tau} v_{\alpha\lambda}(t_1 -\tau  ,t_2 - \tau) \,.
\ee 
The constant $\lambda$ scales away, as it must or else there would be too many solutions, and $v_{\alpha \omega}$ becomes,
\be \label{intC}
 v_{\alpha\omega}(t_1,t_2) = \frac{\sgn(t_{12})}{|t_{12}|^{2 \alpha}} \int_{-\infty}^{\infty} d \tau e^{- i\omega \tau} |1-t_1 +\tau|^{2\alpha-3/2} |1-t_2 + \tau|^{2\alpha-3/2}~.
\ee
Splitting the integral into three regions, depending on how $\tau$ compares with $1-t_1$ and $1-t_2$, and recalling the integral definition of the Bessel functions,
\begin{eqnarray}
K_{\nu} (z) &=&  \frac{(2 z)^{\nu} \sqrt{\pi}}{\Gamma(\nu + 1/2)} e^{-z} \int_0^{\infty} dt\ e^{-2 z t} t^{\nu - 1/2} (1+t)^{\nu - 1/2}~,\\
I_{\nu}(z) &=& \frac{(2z)^{\nu}}{\sqrt{\pi}\Gamma(\nu+1/2)}  e^{z} \int_0^1 dt\ e^{- 2 z t}\(t(1-t)\)^{\nu-1/2}~,
\end{eqnarray}
we find,
\be \label{eginSL0}
 v_{\alpha\omega}(t_1,t_2)  
= \frac{\sgn (t_{12})}{|t_{12}|} e^{-i\omega(t_1+t_2)/2}  \Big(\cos(2\pi \alpha) J_{1-2\alpha}(|\omega t_{12}|/2) + (1+\sin(2\pi \alpha)) J_{2\alpha-1}(|\omega t_{12}|/2) \Big)~,
\ee
where in going from (\ref{intC}) to (\ref{eginSL0}) we have dropped overall factors.\footnote{It is important that in (\ref{eginSL0}) the argument of the Bessel function has $|\omega|$ rather than $\omega$. We are grateful to J.~Maldacena and D.~Stanford for noticing this error in the draft through comparison with their four-point function results \cite{MStoappear}.}
\subsection{Directly finding the eigenvectors} \label{sec:direct}
In the previous section, we used the $SL(2,\mathcal{R})$ symmetry to find the eigenvectors (\ref{eginSL0}). It is useful to directly check that (\ref{eginSL0}) are in fact eigenvectors of the kernel, which is what we do in this section. Aside from being a consistency check, this will also help  to establish for  which choices of $\alpha$ the claimed eigenvectors in (\ref{eginSL0}) are in fact eigenvectors. 

We take the eigenvectors to be of the form,
\be \label{eigTrial}
v_{\nu \omega} (t_a, t_b) =  \frac{\sgn(t_a-t_b)}{|t_a - t_b|} e^{- i \omega (t_a +t_b)/2} Z_{\nu}(|\omega (t_a - t_b)|/2)~,
\ee
where at this stage $Z_{\nu}(\omega |t_a - t_b|/2)$ is taken to be an arbitrary function. We will now insert (\ref{eigTrial}) into the eigenvector equation (\ref{eq:eig}) and perform the integral over $t_a +t_b$. The integral appearing in (\ref{eq:eig}) is,
\be \label{Kint0}
\int d t_a \int d t_b \frac{\sgn(t_a-t_b)}{|t_a - t_b|} e^{- i \omega (t_a +t_b)/2} Z_{\nu}(|\omega (t_a - t_b)|/2) \frac{\sgn(t_a - t_3)}{|t_a - t_3|^{1/2}} \frac{\sgn(t_b - t_4)}{|t_b - t_4|^{1/2}}~.
\ee
We let $t= t_a - t_b$, $t_+ = t_a + t_b$, $\tau=t_3-t_4$, $\tau_+ = t_3 +t_4$, and take $\tau>0$. This transforms the integral into
\be
\int dt  \frac{\sgn(t)}{|t|}  \ Z_{\nu}(|\omega t|/2) \int dt_+\ e^{- i \omega t_+/2}\ \frac{\sgn(t_+ - \tau_+ -\tau+t) \sgn(t_+ - \tau_+ + \tau-t)}{\sqrt{|t+t_+ - \tau-\tau_+||t_+-t-\tau_+ +\tau|}}~.
\ee
For the $t_+$ integral we change variables to $\tilde{t}_+ = \frac{t_+ - \tau_+}{|t-\tau|}$ giving, 
\be
e^{- i \omega \tau_+/2}\int dt  \frac{\sgn(t)}{|t|} \ Z_{\nu}(|\omega t|/2)  \int d\tilde{t}_+\ e^{- i \omega |t-\tau| \tilde{t}_+/2}\ \frac{\sgn(\tilde{t}_+-1) \sgn(\tilde{t}_+ + 1)}{|\tilde{t}_+^2 -1|^{1/2}}~.
\ee
Splitting the $\tilde{t}_+$ integral into three regions and evaluating gives
\be \label{int3}
 -\pi e^{- i \omega \tau_+/2}\int dt  \frac{\sgn(t)}{|t|} \ Z_{\nu}(|\omega t|/2)\ \Big(J_0(|\omega (t-\tau)|/2)+Y_0(|\omega (t-\tau)|/2)\Big)~.
\ee
Since the eigenfunctions of the $SL(2,\mathcal{R})$ Casimir (\ref{eq:casimir}) are Bessel functions, the function $Z_{\nu}$ should be some combination of Bessel functions. While any Bessel function is an eigenfunction of the Casimir, as a result of possible boundary terms (as discussed in Appendix~\ref{appendixB}),   it is only for eigenvectors that are an appropriate combination of Bessel functions that the kernel actually commutes with the $SL(2,\mathcal{R})$ generators. In addition, inspection of the Bessel addition formula (\ref{Badd}) also suggests $Z_{\nu}$ is composed of Bessel functions. In any case, using the hint that the $Z_{\nu}$ are composed of Bessel functions, we take the $Z_{\nu}$ to be some combination of Bessel functions $J_{\nu}$ and $J_{-\nu}$ and fix the relative coefficient so as to ensure it is an eigenvector. In Appendix.~\ref{appendixD} we evaluate the integral (\ref{int3}) and find that the $Z_{\nu}$ in the eigenfunctions are given by,
\be \label{eigenF}
Z_{\nu}(x) = J_{\nu}(x) + \xi_{\nu} J_{- \nu}(x), \ \ \ \ \ \ \xi_{\nu} = \frac{\tan (\nu \pi/2) + 1}{\tan( \nu \pi/2) - 1}~,
\ee
and that the corresponding eigenvalues for the eigenvectors (\ref{eigTrial}) are,
\be \label{eigenvalues}
g(\nu) = -\frac{3}{2\nu} \tan\frac{ \nu \pi}{2}~.
\ee
Setting $\nu = 2\alpha-1$ gives back (\ref{eq:g}).
Moreover, the eigenfunctions (\ref{eigTrial}, \ref{eigenF}) agree with the eigenfunctions (\ref{eginSL0}) found previously through use of the $SL(2,\mathcal{R})$ generators. 

\subsection{A complete set of eigenvectors} \label{complete}
In the previous section we established that the eigenvectors of the kernel (\ref{eq:eig}) are given by (\ref{eigTrial}, \ref{eigenF}). In this section, we find the appropriate range of $\nu$ so as to have a set of eigenvectors $v_{\nu \omega}(t_1, t_2)$ that form a complete basis over $t_1, t_2$.\footnote{The eigenvectors have the antisymmetry $v_{\nu \omega}(t_1, t_2) = - v_{\nu \omega}(t_2, t_1)$, consistent with the antisymmetry of fermions. So, the eigenvectors will form a complete basis for antisymmetric functions of $t_1, t_2$. }  We will do this by appealing to the standard fact in quantum mechanics that the full set of continuous and bound states forms a complete basis.

We start with the Bessel equation, 
\be
t^2 J_{\nu}''(t) + t J_{\nu}'(t) + (t^2 - \nu^2) J(t) = 0 ~,
\ee
which, upon substituting $ x= \log t$ becomes, 
\be
-\frac{d^2 J_{\nu}(x)}{d x^2}- e^{2 x} J_{\nu}(x) = - \nu^2 J_{\nu}(x)~. 
\ee
The Bessel equation looks like a Schr{\"o}dinger equation in a potential
\be
V(x) = - e^{2x}~,
\ee
with an energy of $-\nu^2$.

Now notice that  the eigenfunction (\ref{eigenF}) has a term $J_{-\nu}(|\omega t_{12}|/2)$, which diverges at small $|t_{12}|$ for $\text{Re}(\nu)>0$. In terms of the $x$ coordinate, this is a divergence at large negative $x$ for the states with negative energy.~\footnote{In fact, recall that in the calculation of Appendix \ref{appendixD}, to demonstrate that (\ref{eigenF}) is an eigenfunction requires that $|\text{Re}(\nu)| < 1$, as otherwise various integral identities involving $J_{-\nu}$ are not valid. If, however, $\xi_{\nu}$ is chosen to vanish, so that $J_{-\nu}$ is absent, then the eigenfunctions are valid as long as $\text{Re}(\nu)>-1$. }   We get rid of these states by placing boundary conditions at large positive $x$ that force these solutions to vanish. With the presence of these boundary conditions, we have a quantum mechanics problem in a potential that has both bound states and scattering states. The bound states are characterized by $\nu = 3/2 + 2n$ for nonnegative integer $n$, as these are the only choices of $\nu$ that force  $\xi_{\nu}$ (the coefficient of $J_{-\nu}$) to vanish. The scattering states are given by $\nu = i r$ with $0<r<\infty$. 
The complete set of eigenfunctions is therefore given by $\nu = i r$ with $0<r<\infty$ and $\nu = 3/2 + 2n$ for nonnegative integer $n$. \footnote{In fact, this set of choices of $\nu$ was known to Kitaev to be a complete set, based on considerations of the representation theory of $SL(2,\mathcal{R})$ \cite{Kcom}.}

This mixing of UV and IR is strange, but it is a recurrent theme in AdS$_2$/CFT$_1$. It will appear again when we find in Section \ref{sec:4pt} that there is a divergent piece in the four-point function which must be regulated by physics coming from the UV. 

\subsubsection*{Normalization }
Finally, we need to normalize the eigenfunctions. It will be helpful to use the following indefinite integral Bessel identity, 
\be  \label{JJ}
\int \frac{dt}{t} J_{\nu} J_{\mu} = t\frac{ J_{\nu-1} J_{\mu}  -  J_{\nu} J_{\mu-1}}{\nu^2-\mu^2} - \frac{J_{\nu} J_{\mu}}{\mu+\nu}~.
\ee
First, we normalize the discrete eigenfunctions, with $\nu_n = 3/2 + 2n$. 
For $\text{Re}(\alpha), \text{Re}(\beta)>0$,
\be \label{eq:Jorth}
\int_0^{\infty} \frac{d t}{t} J_{\alpha} J_{\beta} = \frac{2}{\pi} \frac{\sin\Big(\frac{\pi}{2}(\alpha-\beta)\Big)}{\alpha^2 - \beta^2}~.
\ee
This is the contribution at $t=\infty$ of the right-hand side of (\ref{JJ}); the piece at $t=0$ vanishes for  $\text{Re}(\alpha), \text{Re}(\beta)>0$.
Thus, 
\be
\int_0^{\infty} \frac{d t}{t} J_{\nu_n} J_{\nu_m}= \frac{\delta_{n,m}}{2\nu_n}~.
\ee
Now consider the continuous ones,  $Z_{\nu}$ with $\nu = i r$ (\ref{eigenF}). This set of eigenfunctions will be delta-function normalizable.  Computing their inner product
\be
\int_0^{\infty} \frac{d t}{t} Z_{i r}^* Z_{i s}~,
\ee
by using (\ref{JJ}) and evaluating at $t=0$ and $t=\infty$, we find that the contribution at $t=\infty$ vanishes. We regulate the contribution from $t=0$ by evaluating it at $t=e^{-1/\epsilon}$. Noting that 
\be
J_{\nu}(e^{-1/\epsilon}) = \frac{e^{-\nu/\epsilon}}{2^{\nu} \Gamma(\nu+1)}+\cdots
\ee
as well as
\be
\delta(x) = \text{lim}_{\epsilon \rightarrow 0} \frac{\sin(x/\epsilon)}{\pi x}~,
\ee
we find,
\be \label{ZZZ}
\int_0^{\infty} \frac{d t}{t} Z_{i r}^* Z_{i s} =  \frac{2\pi}{\Gamma(1+i r)\Gamma(1-i r)}\delta(r-s)=
 2 \frac{\sinh \pi r }{r} \delta(r-s)~,
\ee
where we have used $\xi_{ir}^* \xi_{ir} = 1$ and dropped terms that oscillate rapidly as $\epsilon \rightarrow 0$. 

\subsubsection*{Summary} \label{sec:summary}
We have found that the complete set of eigenfunctions are,
\be \label{eq:eigFinal}
 v_{\nu \omega} (t_a, t_b) =  \frac{\sgn(t_a-t_b)}{\sqrt{4\pi} |t_a - t_b|} e^{- i \omega (t_a +t_b)/2} \left(J_{\nu}(|\omega (t_a - t_b) |/2) + \frac{\tan (\nu \pi/2) + 1}{\tan (\nu \pi/2) - 1} J_{- \nu}(|\omega (t_a - t_b)|/2)\right)~,
\ee
where $\nu$ pure imaginary, $\nu = i r$ with $r>0$, make up the continuous family, and  $\nu = 3/2+ 2 n$ with integer $n\geq 0$ make up the discrete family.
The eigenfunctions have an inner product
\be
(v_{\nu \omega}, v_{\nu' \omega'}) \equiv \int_0^{\infty} |t_1-t_2|\ d |t_1-t_2|\ \int_{-\infty}^{\infty} d(t_1+t_2)\ v_{\nu \omega}^* v_{\nu' \omega'} = N_{\nu} \delta(\nu - \nu') \delta(\omega-\omega')
\ee
where $\delta(\nu - \nu')$ denotes the Kronecker $\delta_{n m}$ for $\nu$ discrete, and the Dirac $\delta(r-s)$ for $\nu$ continuous, and 
\be \label{norm}
N_{\nu} =
\begin{cases}
(2 \nu)^{-1} & \text{for } \nu = 3/2+2n\\
2 \nu^{-1} \sin\pi \nu & \text{for } \nu=i r~.
\end{cases}
\ee

\section{Four-point function} \label{sec:4pt}
Equipped with the eigenvectors of the kernel, we now find the four-point function (\ref{eq:4chi}). 
Since the eigenvectors $v_{\nu \omega}(t_3 ,t_4)$ (\ref{eq:eigFinal}) form a complete set, we can expand the four-point function in terms of them, 
\be
\Gamma (t_1,t_2,t_3,t_4) = \int d\nu d\omega\, \gamma_{\nu\omega}(t_1, t_2)  v_{\nu\omega}(t_3, t_4) ~,
\ee
where the integral over $\nu$ denotes an integral over the imaginary $\nu$ ($\nu = i r, \ r>0$) and a sum over the discrete real $\nu$ ($\nu_n = 3/2+ 2 n, \ n\geq0$), and there are some coefficients $\gamma_{\nu \omega}$. We may similarly expand $\Gamma_0(t_1,t_2,t_3,t_4)$ appearing in the Schwinger-Dyson equation (\ref{KGamma0}),
\be
\Gamma_0(t_1,t_2,t_3,t_4) = \int d\nu d\omega\, \gamma_{0\nu\omega}(t_1, t_2)  v_{\nu\omega}(t_3, t_4) ~.
\ee
Recalling $\Gamma_0$ in (\ref{eq:Gamma}) we get for $\gamma_{0 \nu \omega}$, 
\be
\gamma_{0 \nu \omega}(t_1, t_2) = \frac{3 J }{\sqrt{4 \pi} N_{\nu }} v_{\nu \omega}^*(t_1, t_2)~,
\ee
where $N_{\nu}$ is the normalization factor (\ref{norm}) for the eigenvectors. From the Schwinger-Dyson equation (\ref{eq:Gamma}) we therefore have,
\be \label{eq:4ptSum}
\Gamma (t_1, t_2, t_3, t_4) = \frac{3 J}{\sqrt{4 \pi}}\int d\nu d\omega \frac{v_{\nu \omega}^*(t_1,t_2)\ v_{\nu \omega}(t_3, t_4)}{(1- g(\nu)) N_{\nu}}~,
\ee
where $g(\nu)$ are the eigenvalues (\ref{eigenvalues}). This is our result for the four-point function. The integral over $\nu$ is to signify an integral over the imaginary $\nu$ and a sum over the discrete $\nu = 3/2+2n$. Eq.~\ref{eq:4ptSum} can be viewed as expressing the four-point function as a sum over all intermediate two-particle states. It is reminiscent of a conformal block decomposition. 

If one sets the eigenvalue $g(\nu)=1$, then the eigenvector equation (\ref{eq:eig}) turns into the Bethe-Salpeter equation for two-particle bound states. This is not the decomposition of the four-point function used in (\ref{eq:4ptSum}).  
However, the eigenvectors $v(t_a, t_b)$ of the kernel  for general eigenvalue $g(\nu)$ are perhaps also in themselves of  interest.

There are three pieces appearing in (\ref{eq:4ptSum}). The first is a pole occurring at $\nu = 3/2$, as a result of $g(3/2) =1$. The second is a sum over the remaining discrete $\nu$, $\nu_n = 3/2 + 2n$ with $n\geq 1$. The third is an integral over $\nu = i r$ with $r>0$. 

\subsection*{Divergence}
Since $g(3/2) = 1$, the four-point function (\ref{eq:4ptSum}) is divergent.   
The true four-point function should be finite, so this divergence must be an artifact of taking the IR limit. Indeed, in computations of the four-point function, we made use of the two-point function given by (dropping overall constants) (\ref{eq:2pt}) ,
\be \label{G2}
G(t) = \frac{\sgn(t)}{\sqrt{J |t|}}~,
\ee
and used this for integrals ranging over all times. However, (\ref{G2}) is only valid in the IR: for time separations $J t \gg 1$, as is clear from the derivation of (\ref{eq:2pt}) in going from (\ref{eq:G}) to (\ref{GSig}). 

To see how the divergence goes away in the full theory, note that the two-point function in the UV is given by $\sgn(t)$ (\ref{G0}), and so the true two-point function interpolates between this and the IR form (\ref{G2}). An example of such a function is
\be \label{Gfull}
G(t) = \frac{\sgn(t) }{\sqrt{|J t| +1}}~,
\ee
though of course (\ref{Gfull}) is not the real two-point function; for this one must  actually solve the Schwinger-Dyson equation (\ref{eq:Sig}, \ref{eq:G}) for all $J t$.~\footnote{In fact, a careful study of the Schwinger-Dyson equation shows that the first subleading term one may naively expect is absent. We thank D.~Stanford for sharing his result with us.} A correction to the form of the IR two-point function (\ref{G2}) is already enough to remove the divergence in the four-point function. For instance,  Taylor expanding (\ref{Gfull}) about $J t\gg1$ gives,
\be \label{eq:Gtaylor}
G(t) = \frac{\sgn(t)}{\sqrt{J |t|}}\( 1- \frac{1}{2 J |t|} + \ldots \)~.
\ee
One can then use first order perturbation theory in quantum mechanics, regarding the kernel as a Hamiltonian, to compute the change in the eigenvalues, under the change  $\delta G$ in $G$ going from (\ref{G2}) to (\ref{eq:Gtaylor}), 
\be
\delta g_{\nu} \delta(\omega - \omega') = \int d t_1 d t_2 d t_3 d t_4\ v^*_{\nu \omega'}(t_1,t_2)\ \delta K\ v_{\nu \omega}(t_3,t_4)~.
\ee
The shift in the eigenvalue $\delta g_{\nu}$ will be a power of $\omega/J$ (depending on the power of $(J t)^{-1}$ used in $\delta G$). This shift $\delta g_{\nu}$  removes the divergence of the four-point function and, in addition, removes the degeneracy of the eigenvalues by having them acquire $\omega$ dependance. \footnote{The connection between $\nu=3/2$ and  breaking of conformal symmetry was recognized by Kitaev~\cite{Kitaev}.}

The need to break conformal symmetry is in some sense surprising. In the spirit of effective field theory, one may have thought that the conformal IR theory should in itself be consistent.
Instead, we see that the UV does not truly decouple. 
In fact, this behavior should have been expected from holographic studies in AdS$_2$/CFT$_1$. Gravity in AdS$_2$ is known to be problematic, as the backreaction of any finite energy excitation is so strong that it destroys the boundary \cite{Maldacena:1998uz}. 
This was studied in \cite{AP} by embedding  AdS$_2$ in a higher dimensional space: with Poincare coordinate $z=0$ denoting the boundary, there was a transition at $z=a$ from conformal Lifshitz (small $z$) to $AdS_2$ times a compact space (large $z$). On the boundary, this corresponds to an RG flow with a CFT$_1$ in the IR. From bulk computations, \cite{AP} found breaking of conformal invariance in the four-point function, along with a divergence  as the regulator $a$ was removed. 
 
 This suggests that it is not the IR fixed point of SYK that should be thought of as dual to AdS$_2$. Rather, one should consider an AdS$_2$ embedded in a higher dimensional space: for instance, as  the near-horizon limit of an extremal charged Reisnner-Nordstr{\" o}m black hole in asymptotic AdS. While the dual of this bulk is certainly not SYK, it may be that the IR limit is SYK.

\subsection*{Chaos}\label{sec:chaos}
Building on semiclassical intuition, quantum chaos can be diagnosed by the exponential growth of an out-of-time-order four-point function \cite{LO, KitaevNov}. For recent work, see \cite{Maldacena:2015waa,Roberts:2014ifa, Roberts:2014isa, Hosura:2015ylk, Gur-Ari:2015rcq, Stanford:2015owe, Polchinski:2015cea}. In the context of SYK, one can consider the thermal out-of-time-order four-point function, 
\be \label{OFTO}
\langle \chi_i(0)\chi_j(t)\chi_i(0) \chi_j(t)\rangle_{\beta}~ \sim \frac{1}{N} e^{\kappa t}~,
\ee
where $\kappa$ is identified with the Lyapunov exponent.\footnote{Eq.~\ref{OFTO} is valid for times longer than the dissipation time and shorter than the scrambling time, $\kappa^{-1} \ll t\ll \kappa^{-1} \log N$.} Kitaev found that SYK, at strong coupling $\beta J \gg 1$, has a Lyapunov exponent $\kappa = 2\pi/\beta$ \cite{Kitaev}. This was done by considering the Schwinger-Dyson equation for the out-of-time-order four-point function, as defined on the Keldysh contour, and plugging in an ansatz of the form (\ref{OFTO}). 

One can also compute (\ref{OFTO}) from the zero-temperature Euclidean four-point function (\ref{eq:4chi}). 
We begin by noting that the finite temperature four-point function can be obtained by a conformal mapping of the zero-temperature four-point function. The Schwinger-Dyson equation for the two-point function (\ref{eq:Int2pt2}) had the invariance (\ref{Gconf}), $G'(t_1, t_2) = \[\partial_1 f(t_1) \partial_2 f(t_2)\]^{1/4} G(f(t_1), f(t_2))$. Similarly, one can check that if  $\Gamma(t_1, t_2, t_3, t_4)$ satisfies  the Schwinger-Dyson equation for the four-point function (\ref{eq:Gamma}), then so does
\be \label{GammaConf}
\Gamma'(t_1, t_2, t_3, t_4) = \[ \partial_1 f(t_1) \partial_2 f(t_2) \partial_3 f(t_3) \partial_4 f(t_4)\]^{3/4} \Gamma(f(t_1), f(t_2), f(t_3), f(t_4))~,
\ee
while the eigenvectors transform as,
\be
v'_{\nu \omega} (t_1, t_2) = \[\partial_1 f(t_1) \partial_2 f(t_2)\]^{3/4} v_{\nu \omega}(f(t_1), f(t_2))~.
\ee
The finite temperature four-point function therefore follows from the zero-temperature one through the mapping $f(t) =\exp(2 \pi i t/\beta)$. 

Finally, the Euclidean correlator is  transformed into a Lorentzian out-of-time-order correlator by assigning small Euclidean time $\epsilon_j$ to $t_j$ (the desired ordering of the Lorentzian correlator determines the relative magnitude of the $\epsilon_j$), then adding some Lorentzian time, and finally sending $\epsilon_j$ to zero  (see e.g\  \cite{Haag, Roberts:2014ifa, Hartman:2015lfa}). In particular, for (\ref{OFTO}) one chooses,
\be \label{eq:f}
f(t_1)  = e^{\frac{2 \pi}{\beta} i \epsilon}\ ,\,  \, \, \, f(t_3) = e^{\frac{2 \pi}{\beta} 2 i \epsilon} e^{2 \pi t/\beta}~,\, \, \, \, f(t_2) = e^{\frac{2 \pi}{\beta} 3 i \epsilon}~,\, \, \, \, f(t_4) = e^{\frac{2 \pi}{\beta} 4i \epsilon} e^{2 \pi t/\beta}~.
\ee

\subsection*{Discrete Sum} \label{sec:sum}
We now  return to the expression for the four-point function, (\ref{eq:4ptSum}), and evaluate  the sum over the discrete $\nu_n = 3/2+2n$ with $n\geq1$. Denoting this by $\Gamma^d$, we get from (\ref{eq:4ptSum}), 
\begin{multline} \label{Gamma24}
\Gamma^d (t_1, t_2, t_3, t_4) \\
=\frac{3J }{ \pi^{3/2}} \frac{\sgn(t_{12})\sgn(t_{34})}{|t_{12}| |t_{34}| }\sum_{n=1}^{\infty}\frac{(n+3/4)^2}{n} \int_0^{\infty} d\omega\ \cos (\omega s/2)\ J_{3/2+2n}(\omega |t_{12}|/2) J_{3/2+2n}(\omega |t_{34}|/2) ~,
\end{multline}
where we have defined,
\be \label{eq:s}
s\equiv t_3+t_4-t_1-t_2~.
\ee
Using Eq.~6.612 of \cite{Grad} (see Appendix~\ref{appendixE}), one has that,
\be  \label{IntQ}
\int_0^{\infty} d x\  J_{\nu}(a x) J_{\nu}(b x) \cos( s x)\\ = \frac{1}{2}\[Q_{\nu-1/2}\Big(\frac{b^2 +a^2 - s^2}{2 a b} +i \epsilon \Big)+Q_{\nu-1/2}\Big(\frac{b^2 +a^2 - s^2}{2 a b} -i \epsilon \Big)\]~.
\ee
The Legendre function of the second kind, $Q_{\nu}(z)$, has a branch cut along a portion of the real axis, $z\in (-\infty, 1)$. As discussed in Appendix~\ref{appendixE}, we define $Q_{\nu}(z)$ through a hypergeometric function (\ref{QF}). For arguments $z>1$, one has that $\frac{1}{2}(Q_{\nu}(z+i\epsilon)+ Q_{\nu}(z-i\epsilon))= Q_{\nu}(z)$. For $z<-1$, one finds that $\frac{1}{2}(Q_{\nu}(z+i\epsilon)+ Q_{\nu}(z-i\epsilon) )= -\cos (\nu \pi) Q_{\nu}(z)$.   (see Eq. \ref{QcosI}). An alternative way to define $Q_{n}(z)$ is as a Hilbert transform of the Legendre function of the first kind, $P_{n}(z)$, 
\be \label{eq:QH}
Q_{n}(x) = \frac{1}{2}\int_{-1}^{1} dz \frac{P_{n}(z)}{x-z}~,
\ee
where for $x\in(-1,1)$, the integral is interpreted as a Cauchy principal value integral.  For $x\in(-1,1)$, the definition (\ref{eq:QH}) of $Q_{\nu}(z)$ is equal to $\frac{1}{2}(Q_{\nu}(z+i\epsilon)+Q_{\nu}(z-\epsilon))$, where in the latter $Q_{\nu}$ denotes the definition through the hypergeometric function (\ref{QF}). 
With this understanding, we have from (\ref{Gamma24}) and (\ref{IntQ}),
\be \label{GammadS}
\Gamma^d (t_1, t_2, t_3, t_4) = \\
\frac{ 3 J }{\pi^{3/2} } \frac{\sgn(t_{12}) \sgn(t_{34})}{|t_{12}|^{3/2} |t_{34}|^{3/2}}\sum_{n=1}^{\infty}\frac{(n+3/4)^2}{n} 
Q_{2n+1}\Big(\frac{ t_{12}^2 +t_{34}^2 - s^2}{2 |t_{12}| |t_{34}|}\Big)~.
\ee
We have written (\ref{GammadS}) for the case when the argument of the Legendre function $Q_{2n+1}(x)$ is in the range $x\in (-1,\infty)$ (where for $x\in(-1,1)$ the definition (\ref{eq:QH}) is implied). For $x\in (-\infty, -1)$, one should replace $Q_{2n+1}(x)$ with $Q_{2n+1}(-x)$. 

We note that the first several Legendre functions are given by,~\footnote{This is for $x\in (-1,1)$. For $x>1$, the argument of the $\log$ should get an extra minus sign.}
\begin{eqnarray}  \nonumber
P_0(x) &=& 1 \ \ \ \ \ \ \ \ \ \ \ \ \ \ \ \ \ \ \ \ \ \ \ \   Q_0(x) = \frac{1}{2}\log \frac{1+x}{1-x} \\ \nonumber
P_1(x) &=& x\ \ \ \ \ \ \ \ \ \ \ \ \ \ \ \ \ \ \ \ \ \ \ \  Q_1(x) = -1+ \frac{x}{2} \log \frac{1+x}{1-x} \\ \nonumber
P_2(x) &=& \frac{1}{2}( 3 x^2-1) \ \ \ \ \ \ \ \ \ \  Q_2(x)  = -\frac{3 x}{2}  +\frac{1}{4}(3x^2 - 1)\log \frac{1+x}{1-x}~.
\end{eqnarray}

The form of the discrete tower contribution to the four-point function, expressed as the sum (\ref{GammadS}), is already an interesting expression and should be studied further.~\footnote{For instance, one can study the themal out-of-time-order four-point function, through the conformal mapping specified in (\ref{GammaConf}) and (\ref{eq:f}). The contribution of the $\nu_0=3/2$ block (not included in the sum, since it is the one with a divergent coefficient) gives exponential growth with the exponent $ 2\pi/\beta$, while simply from the leading  scaling of $Q_{2n+1}$, one can anticipate that the $\nu_n=3/2+2n$ block gives an exponent that has an additional factor of $2n+1$ (of course, the overly rapid growth of the individual terms is not a problem, as the full sum over all $n\geq 1$ cures this).}
We will now evaluate the sum over $n$ in (\ref{GammadS}). We first evaluate the same sum as in (\ref{GammadS}), but with the $P_{2n+1}$ instead of the $Q_{2n+1}$. Recall that the Legendre functions $P_{n}(x)$ can be found from the generating function $h(v,x)$,~\footnote{The $Q_n(x)$ also have a generating function, which is just the Hilbert transform of $h(v,x)$.}
\be \label{eq:h}
h(v,x) = \frac{1}{\sqrt{1 - 2 x v + v^2}} \equiv \sum_{k=0}^{\infty} P_k(x) v^k~.
\ee
We will evaluate our sum by taking derivatives and integrals of the generating function, so as to appropriately form the rational function of $n$ appearing in the sum.
Letting
\be
\bar{H}(v,x) = \int dv \frac{h(v,x)}{v^2} 
\ee
so that
\be
H(v,x)\equiv \bar{H}(v,x) - P_1(x) \log v + \frac{P_0(x)}{v} = \sum_{k=2}^{\infty} P_k(x) \frac{v^{k-1}}{k-1}
\ee
and 
\be
h_{2}(v,x) \equiv v^{1/2} \partial_v\( v \partial_v(v^{3/2} H(v,x))\) = \sum_{k=2}^{\infty} P_k(x) \frac{(k+1/2)^2}{k-1} v^k
\ee
and
\be
h_3(x) \equiv h_2(1,x) - h_2(-1,x)~,
\ee
We thus get,
\be \label{eq:h3Sum}
h_3(x) =4 \sum_{n=1}^{\infty}\frac{(n+3/4)^2}{n} P_{2n+1}(x)~.
\ee
where performing the above operations we find,
\be \label{h3}
h_3(x) =-6 x+ \frac{3}{\sqrt{2}} \frac{ (1+3x/2)}{\sqrt{1+x}}- \frac{3}{\sqrt{2}} \frac{ (1-3x/2)}{\sqrt{1-x}}- \frac{9}{4}x \log\((1-x+\sqrt{2}\sqrt{1-x} )(1+x+\sqrt{2}\sqrt{1+x}) \)~.
\ee
Since the Legendre $Q_n$ are defined in terms of $P_n$ by a Hilbert transform (\ref{eq:QH}), we can get the sum
\be
\tilde{h}_3(x) =4 \sum_{n=1}^{\infty}\frac{(n+3/4)^2}{n} Q_{2n+1}(x)~,
\ee
by performing a Hilbert transform of (\ref{h3}), 
\be \label{tildh}
\tilde{h}_3(x) = \frac{1}{2}\int_{-1}^1 dz \frac{h_3(z)}{x-z}~.
\ee
This Hilbert transform is straightforward to evaluate and contains, for instance, the dilogarithm function. We thus finally get,
\be \label{Gammad}
\Gamma^d (t_1, t_2, t_3, t_4) = 
\frac{3 J }{4\pi^{3/2} }\frac{\sgn(t_{12})}{|t_{12}|^{3/2}}\frac{\sgn(t_{34})}{ |t_{34}|^{3/2}}  \tilde{h}_3\Big(\frac{ t_{12}^2 +t_{34}^2 - s^2}{2 |t_{12}| |t_{34}|}\Big) ~.
\ee


\subsubsection*{Continuous Sum}
Finally, the four-point function $\Gamma$ (\ref{eq:4ptSum}) also has a contribution coming from an integral over $\nu = i r$. Denoting this piece by $\Gamma^c$, 
and inserting the eigenfunctions (\ref{eq:eigFinal}), eigenvalues (\ref{eigenvalues}), and normalization (\ref{norm}) into (\ref{eq:4ptSum}), 
\be \label{eq:GammaC} \nonumber
\Gamma^c(t_1,t_2,t_3,t_4) = \frac{3 J}{(4\pi)^{3/2}}\frac{1}{ t_{12} t_{34}} \int_0^{\infty} dr \int_{-\infty}^{\infty} d\omega\ r^2 e^{- i \omega s/2}\ \frac{ Z_{i r}^*(|\omega t_{12}|/2)\ Z_{ir}(|\omega t_{34}|/2)}{2 r \sinh(\pi r) + 3 \cosh(\pi r) -3}~,
\ee
where $Z_{\nu}$ is given by (\ref{eigenF}) and $s$ is defined by (\ref{eq:s}). In Appendix~\ref{appendixE} we perform the integral over $\omega$; the remaining integral over $r$ is left to future work.

\acknowledgments
We thank O.~Aharony, D.~Gross, A.~Kitaev, Z.~Komargodski, J.~Maldacena, B.~Shraiman, D.~Stanford  for helpful discussions, and thank A.~Kitaev, J.~Maldacena, D.~Stanford for comments on the draft. This work was supported by  NSF Grants PHY11-25915 and PHY13-16748. 

\appendix

\section{Integrals in eigenvalue computation} \label{appendixA}
In this appendix we evaluate the integral appearing in the computation of the eigenvalues of the kernel in Sec.~\ref{sec:eigenvalues1}. 

We will need to evaluate the integral on the left-hand side of (\ref{eq:eig}),
\be
\int d t_a \int d t_b \frac{ \sgn(t_a - t_b)}{|t_a - t_b|^{2 \alpha}}\ \frac{\sgn(t_a - t_3)}{|t_a - t_3|^{2 \Delta}}\ \frac{\sgn(t_b - t_4)}{|t_b - t_4|^{2 \Delta}}~.
\ee
There are $8$ regions of integration which must be done separately. Without loss of generality, we let $t_3>t_4$.
The following representations of the $\beta$ function will be useful,
\be
\beta(x,y) = \frac{ \Gamma(x) \Gamma(y)}{\Gamma(x+y)}
\ee
\be
\int_{-\infty}^{\tau} d t \frac{1}{(s-t)^x} \frac{1}{(\tau - t)^y} = \int_s^{\infty} dt \frac{1}{(t-s)^y}\ \frac{1}{(t-\tau)^{x}}= \frac{1}{(s-\tau)^{x+y-1} }\beta(1-y, x+y-1) 
\ee
\be
\int_s^{\tau} dt \frac{1}{(t-s)^x}\ \frac{1}{(\tau -t)^y} = \frac{1}{(\tau- s)^{x+y -1}} \beta(1-x, 1-y)~.
\ee
\\
We label the ranges by indicating if the $\sgn$ is positive or negative. 

1. $+++$
\begin{multline} \label{eq:int1}
\int _{t_3}^{\infty} d t_a \int _{t_4}^{t_a} d t_b\ \frac{1}{(t_a - t_b)^{2\alpha}}\ \frac{1}{(t_a - t_3)^{2\Delta}}\ \frac{1}{(t_b- t_4)^{2 \Delta}}\\
= \beta(1 - 2\alpha, 1 - 2 \Delta) \int_{t_3}^{\infty} d t_a\ \frac{1}{(t_a-t_3)^{2 \Delta}}\ \frac{1}{(t_a - t_4)^{2 \alpha + 2 \Delta -1}}\\
= \beta(1-2\alpha, 1 - 2 \Delta) \beta(1 - 2\Delta, 2\alpha +4 \Delta - 2)\ \frac{1}{(t_3-t_4)^{2 \alpha +4 \Delta -2}}
\end{multline}
From now on we will not write the time dependance of the result of integrals, and use notation $t_{i j} = t_i -t_j$.

2. $++-$
\be
- \int_{-\infty}^{t_4} d t_b \int_{t_3}^{\infty} d t_a \ \frac{1}{t_{ab}^{2\alpha}}\ \frac{1}{t_{a3}^{2\Delta}}\ \frac{1}{t_{4b}^{2 \Delta}} = - \beta(1- 2\Delta, 2 \alpha + 2\Delta -1) \beta(1-2\Delta, 2\alpha +4 \Delta -2)
\ee

3. $+-+$
\be
- \int_{t_4}^{t_3} d t_a\ \int_{t_4}^{t_a} d t_b \frac{1}{t_{a b}^{2\alpha}}\ \frac{1}{t_{3 a}^{2\Delta}}\ \frac{1}{t_{b 4}^{2 \Delta}}
= - \beta(1-2\alpha, 1- 2\Delta) \beta(1-2\Delta, 2 - 2\alpha - 2\Delta)
\ee

4. $+--$
\be
\int_{-\infty}^{t_4} d t_b \ \int_{t_b}^{t_3} dt_a\ \frac{1}{t_{a b}^{2\alpha}}\ \frac{1}{t_{3 a}^{2\Delta}}\ \frac{1}{t_{4 b}^{2 \Delta}}
= \beta(1-2\alpha, 1- 2 \Delta) \beta(1-2 \Delta, 2 \alpha +4 \Delta -2)
\ee

5. $-++$
\be
- \int_{t_3}^{\infty} d t_b\ \int_{t_3}^{t_b} d t_a\ \frac{1}{t_{b a}^{2\alpha}}\ \frac{1}{t_{a 3}^{2\Delta}}\ \frac{1}{t_{b 4} ^{2 \Delta}}
= -\beta(1- 2\alpha, 1 - 2 \Delta) \beta(2-2\alpha -2 \Delta, 2\alpha + 4 \Delta -2)
\ee

6. $-+-$.

\ \ \ \ \ Doesn't exist, since we assumed $t_3>t_4$. 

7. $--+$

We need to split the integral into two regions, 
\be \label{eq:7a}
\int_{-\infty}^{t_4} d t_a\ \int_{t_4}^{\infty} d t_b\ \frac{1}{t_{b a}^{2\alpha}}\ \frac{1}{t_{3 a}^{2\Delta}}\ \frac{1}{t_{b 4}^{2 \Delta}} 
= \beta(1-2\Delta, 2 \alpha +2 \Delta -1) \beta(2-2\alpha - 2\Delta, 2\alpha +4 \Delta -2)
\ee
and 
\be\label{eq:7b}
\int_{t_4}^{t_3} d t_a\ \int_{t_a}^{\infty} d t_b\ \frac{1}{t_{b a}^{2\alpha}}\ \frac{1}{t_{3 a}^{2\Delta}}\ \frac{1}{t_{b 4}^{2 \Delta}} 
= \beta(1-2\alpha, 2 \alpha + 2\Delta -1) \beta(1-2\Delta, 2- 2\alpha - 2\Delta)
\ee

8. $---$
\be \label{eq:int8}
-\int_{-\infty}^{t_4} d t_a \int_{t_a}^{t_4} d t_b  \frac{1}{t_{b a}^{2\alpha}}\ \frac{1}{t_{3 a}^{2\Delta}}\ \frac{1}{t_{4 b}^{2 \Delta}} 
= - \beta(1-2\alpha, 1-2\Delta) \beta(2-2\alpha - 2\Delta, 2 \alpha +4 \Delta -2)
\ee
Summing the results (\ref{eq:int1}) - (\ref{eq:int8}) and recalling $\Delta = 1/4$ gives $g(\alpha)$,
\be 
g(\alpha) = - \frac{3}{2} \frac{1}{(1- 2 \alpha) \tan (\pi \alpha)}~.
\ee

\section{Eigenvectors and boundary terms} \label{appendixB}
In this appendix, we elaborate on the statement made in Sec.~\ref{sec:sl2} that care must be taken in arguing that the $SL(2,\mathcal{R})$ generators commute with the kernel. 

In particular, to show that $\tilde L_2  v_{\alpha\omega}$ is an eigenvector if $v_{\alpha\omega}$ is an eigenvector, one must integrate by parts
\begin{multline} \label{eq:bdy}
\int dt_a dt_b\ (t_a^2\partial_a + t_b^2\partial_b) \tilde v_{\alpha \omega} K(t_a, t_b, t_3, t_4) = - \int d t_a dt_b\ \tilde v_{\alpha \omega}\Big( \partial_a(t_a^2 K) +\partial_b(t_b^2 K)\Big)\\
+ \int dt_b\  (\tilde v_{\alpha \omega} t_a^2 K) \Big|_{t_a = - \infty}^{t_a=\infty} + \int d t_a \  (\tilde v_{\alpha \omega} t_b^2 K) \Big|_{t_b = - \infty}^{t_b=\infty}
\end{multline}
We will need to drop the boundary term on the second line. As we will see, this assumption will only be true in certain cases. 

To find the eigenvectors, we use the naive commutativity~(\ref{lkkl}) to conclude that the eigenfunctions of the kernel are the same as those of the $SL(2,\mathcal{R})$ Casimir.  The latter is
\be
2 \tilde L^2 = 2 \tilde L_1^2 - L_0 \tilde L_2 - \tilde L_2 L_0~.
\ee
We find
\bea \label{eq:casimir}
\tilde L^2 &=& t_-^2 \partial_-^2 + 3 {t_-} \partial_- - t_-^2 \partial_+^2   \nonumber\\
&=&  t_-^{-3/2}\left(t_-^2 (\partial_-^2 - \partial_+^2) + \frac{3}{4}  \right) t_-^{3/2} \nonumber\\
&=&  t_-^{-1} \left(t_-^2 \partial_-^2 + t_- \partial_- + t_-^2 \omega^2 + 1\right) t_- \,.
\eea
Here $t_\pm = \frac12( t_1 \pm t_2)$.  In the second line we see that the Casimir is the Lorentzian Laplacian, even for Euclidean four-point functions.  In the third, we have gone to frequency space, and we see that the Casimir is conjugate to the Bessel operator, plus a constant.  Thus,
\be 
 \frac{\sgn(t_1-t_2)}{|t_1 - t_2|} e^{- i \omega (t_1 +t_2)/2} J_{2 \alpha - 1}(|\omega (t_1 - t_2) |/2) 
\ee 
is an eigenfunction of $\tilde{L}^2$, and hence would seem to be an eigenvector of the kernel $K$ as well. Note that  $Y_{2 \alpha-1}$ would also seem to be  an eigenvector.
The important point is that to drop the boundary term appearing in (\ref{eq:bdy}) requires a particular combination of the Bessel functions, such that this term actually vanishes. The eigenvectors that we found in the main text formed the correct combination so that this boundary term vanishes.

\section{Integral in eigenvector computation}  \label{appendixD}
In this appendix we perform the integral appearing in the direct calculation of the eigenvectors in Sec.~\ref{sec:direct}. 

We need to evaluate the integral (\ref{int3}),
\be \label{int3a}
 -\pi e^{- i \omega \tau_+/2}\int dt  \frac{\sgn(t)}{|t|} \ Z_{\nu}(|\omega t |/2)\ \Big(J_0(|\omega (t-\tau)|/2)+Y_0(|\omega (t-\tau)|/2)\Big)~.
\ee
We can  rewrite the integral as,
\bea
&& \int_{-\infty}^{\infty} \frac{dt}{t} Z_{\nu}(|t|)\Big(J_0(|\omega \tau/2-t|) + Y_0(|\omega \tau/2-t|)\Big)\\ \label{eq:eignZ}
&=& \int \frac{d p}{2\pi} e^{- i  p \omega \tau /2} \tilde{Z}_{\nu}'(p) \Big(\tilde{J}_0(p) +\tilde{Y}_0(p)\Big)~,
\eea
where 
\bea
\tilde{J}_0(p) &=& \int dt\ e^{i p t}\ J_0(|t|)= \frac{2}{\sqrt{1-p^2}} \theta(1-|p|)\\
\tilde{Y}_0(p) &=&\int dt\ e^{i p t}\ Y_0(|t|)  = - \frac{2}{\sqrt{p^2-1}} \theta(|p|-1)~.
\eea
Also, 
\be
\tilde{J}_{\nu}'(p) = \int \frac{dt}{t}\ e^{i p t} J_{\nu}(|t|)=\frac{2 i\ \sgn(p)}{\nu}\Big( \sin(\nu \sin^{-1}|p|) \theta(1-|p|) + \frac{\sin(\nu \pi/2)}{(|p|+\sqrt{p^2-1})^{\nu}}\theta(|p|-1)\Big)
\ee
and similarly for $Y_{\nu}'(p)$ (Eq.~6.693 of \cite{Grad}). One should note that the above formula for $J_{\nu}'(p)$ is only valid for $\text{Re }  \nu >-1$, and the one for $Y_{\nu}'(p)$ is valid for $|\text{Re}(\nu)|<1$. 

Let the eigenvector be a combination of Bessel functions, 
\be \label{Zstart}
Z_{\nu} = c_J J_{\nu} + c_Y Y_{\nu}~.
\ee
The Fourier transform of (\ref{eq:eignZ}) becomes,
\begin{multline}  \label{FT1}
\frac{4i}{\nu \sqrt{1-p^2}}\theta(1-|p|) \sin(\nu \sin^{-1} |p|) \Big( c_J - c_Y \tan \nu \pi/2\Big) \\
-\frac{2i}{\nu \sqrt{p^2-1}}\theta(|p|-1) (|p| - \sqrt{p^2-1})^{\nu}  \frac{\cos(\nu \pi)}{\cos(\nu \pi/2)} c_Y\\
 -\frac{4i}{\nu \sqrt{p^2-1}}\theta(|p|-1)\Big(\frac{c_J \sin (\nu \pi/2)}{(|p| +\sqrt{p^2-1})^{\nu}} - \frac{1}{2}\frac{c_Y}{\cos(\nu \pi/2) (|p| - \sqrt{p^2-1})^{\nu}}\Big)
\end{multline}
The Fourier transform of $Z_{\nu} \sgn(\tau)$ is 
\begin{multline} \label{FT2}
 \frac{2i \sin(\nu \sin^{-1}|p|)}{\sqrt{1-p^2}}\theta(1-|p|)\Big(c_J + \frac{c_Y}{\tan(\nu \pi/2)}\Big)\\
 + \frac{i \cos(\pi \nu)}{\sin(\nu \pi/2)\sqrt{p^2-1}}(|p| - \sqrt{p^2-1})^{\nu} \theta(|p|-1)   c_Y\\
+  \frac{ i}{\sqrt{p^2-1}} \theta(|p|-1)\Big(2 \frac{c_J  \cos \nu \pi/2 }{ (|p|+\sqrt{p^2-1})^{\nu}} - \frac{c_Y}{\sin(\nu \pi/2)  (|p|-\sqrt{p^2-1})^{\nu}}\Big)~,
\end{multline}
which has the range of validity of $\text{Re}(\nu)>-2$ coming from the $J_{\nu}$ integral, and $|\text{Re } \nu| <2$ from the $Y_{\nu}$ integral. 
Equating (\ref{FT1}) and (\ref{FT2}), the eigenfunction is therefore, 
\be
\bar{Z}_{\nu} = \(\tan \nu \pi/2 - 1\) J_{\nu} + \(1+\tan\nu \pi/2\) Y_{\nu}~,
\ee
with eigenvalues  $ \frac{2\pi}{\nu} \tan \nu \pi/2$ (recall the factor of $-\pi$ in (\ref{int3a})).
We can rewrite this as
\be
Z'_{\nu}\equiv -\bar{Z}_{\nu} \sin{\nu \pi}=\Big( J_{\nu} (\tan \nu \pi/2 - 1) + J_{- \nu}( \tan \nu \pi/2 + 1)\Big)
\ee
Finally, let us rescale the eigenfunctions, writing them as 
\be \label{appZ}
Z_{\nu} = J_{\nu} + \xi_{\nu} J_{- \nu}, \ \ \ \ \ \ \xi_{\nu} = \frac{\tan \nu \pi/2 + 1}{\tan \nu \pi/2 - 1}~,
\ee
where we are reusing notation for $Z_{\nu}$; this $Z_{\nu}$ is a multiple of the one in (\ref{Zstart}). Now recall that in the integral (\ref{Kint0}) there should be a factor of $ - \frac{3}{4\pi}$: the $3$ is due to Feynman diagram combinatorics, and the $1/4\pi$ is from the normalization of the 2-pt function. The eigenvalues are thus,
\be
g(\nu) = -\frac{3}{2\nu} \tan\frac{ \nu \pi}{2}~.
\ee
Setting $\nu = 2\alpha-1$ gives (\ref{eq:g}).
Moreover, the eigenfunctions (\ref{appZ}) agree with (\ref{eginSL0}).
Finally, it will be useful for later to note that
\be
\xi_{ i r} = -\frac{1}{\cosh(\pi r)}( 1 + i\sinh \pi r)~,
\ee
and so $Z_{ir}^* = Z_{- i r}$.

\section{Integrals of products of Bessel functions} \label{appendixE}
In this appendix we review some integral identities involving products of Bessel functions. 

\subsection*{Laplace transform of $J_{\nu} J_{\nu}$ and $J_{\nu} J_{- \nu}$}
We would like to evaluate integrals of the form
\be
\int_0^{\infty}dt\ e^{- \alpha t} J_{\nu}(\beta t) \mathcal{J}_{\nu}(\gamma t)~,
\ee
where the cylindrical function $\mathcal{J}_{\nu}$ is defined as 
\be \label{Jmath}
\mathcal{J}_{\nu} = a(\nu) J_{\nu} + b(\nu) Y_{\nu}~,
\ee
where $a(\nu), b(\nu)$ are arbitrary functions of $\nu$ with period one, and $J_{\nu}, Y_{\nu}$ are the Bessel functions.

From the Bessel addition formula,
\be \label{Badd}
J_{\nu}\(\sqrt{Z^2 +z^2 - 2 Z z \cos \phi}\)\(\frac{Z- z e^{-i \phi}}{Z - ze^{i \phi}}\)^{\nu/2} = \sum_{m=-\infty}^{\infty} J_{\nu+m}(Z) J_m(z) e^{i m \phi}~,
\ee
where $|z e^{\pm i \phi}|<|Z|$, one finds \cite{Watson}, 
\be \
\int_0^{\pi} d \phi \frac{\mathcal{J}_{\nu}(\sqrt{Z^2 +z^2 - 2 Z z \cos \phi})}{(Z^2 +z^2 - 2 Z z \cos \phi)^{\nu/2}} \sin^{2\nu} \phi  = 2^{\nu} \Gamma(\nu +1/2) \Gamma(1/2) \frac{\mathcal{J}_{\nu}(Z)}{Z^{\nu}} \frac{J_{\nu}(z)}{z^{\nu}}~.
\ee
 Next, following the same procedure as  in \cite{Watson}, and defining $\bar{\omega} = \sqrt{\beta^2 + \gamma^2- 2\beta\gamma \cos \phi}$, one has that, 
\be \label{JmJ}
\int_0^{\infty}dt\ e^{- \alpha t} J_{\nu}(\beta t) \mathcal{J}_{\nu}(\gamma t) = \frac{(\frac{1}{2} \beta \gamma)^{\nu}}{\Gamma(\nu + \frac{1}{2}) \Gamma(\frac{1}{2})} \int_0^{\infty} dt \int_0^{\pi} d\phi\ e^{- \alpha t}\ \frac{ \mathcal{J}_{\nu}(\bar{\omega} t)}{{\bar{\omega}}^{\nu}} t^{\nu} \sin^{2 \nu} \phi ~,
\ee
where $|\beta|<|\gamma|$.  Now using Eq.~13-2 (2) of \cite{Watson}:
\be
\int_0^{\infty} d t\ e^{- \alpha t}\ J_{\nu}(\bar{\omega} t) t^{\rho -1} = \frac{(\bar{\omega}/2\alpha)^{\nu}\Gamma(\rho+\nu)}{\alpha^{\rho} \Gamma(\nu +1)}\  \phantom{}_2 F_1 \(\frac{\rho+\nu}{2}, \frac{\rho+\nu+1}{2}, \nu+1, -\frac{\bar{\omega}^2}{\alpha^2}\)~.
\ee
Combing the previous several lines, and taking $\mathcal{J}_{\nu}= J_{\nu}$ in (\ref{JmJ}) gives,
\begin{eqnarray} \label{JJint} \nonumber
\int_0^{\infty}dt\ e^{- \alpha t} J_{\nu}(\beta t) J_{\nu}(\gamma t) &=& \frac{(\beta \gamma)^{\nu}}{\pi \alpha^{2\nu +1}} \int_0^{\pi}d \pi \sin^{2\nu} \phi\  \phantom{}_2 F_1(\nu+1/2, \nu+1, \nu+1, - \frac{\bar{\omega}^2}{\alpha^2})\\ \nonumber
&=& \frac{(\beta \gamma)^{\nu}}{\pi \alpha^{2\nu +1}} \int_0^{\pi}d \phi \sin^{2\nu} \phi\ \(1+\frac{\bar{\omega}^2}{\alpha^2}\)^{-1/2 - \nu}\\\nonumber
&=& \frac{1}{\pi\sqrt{\beta\gamma}}\frac{2^{\nu-1/2}}{(1+z)^{1/2 + \nu}}\frac{\Gamma(1/2 + \nu)^2}{\Gamma(1+2\nu)} \ \phantom{}_2 F_1\(\nu+1/2, \nu+1/2, 2\nu+1,\frac{2}{1+z}\)\\
&=& \frac{1}{\pi\sqrt{\beta\gamma}} Q_{\nu-1/2}(z)~,
\end{eqnarray}
where in the last line we used the relation,
\be
\phantom{}_2 F_1(a, b, 2 b, x) = (1 - x/2)^{-a}\ \phantom{}_2 F_1\(\frac{a}{2}, \frac{a + 1}{2}, b + \frac{1}{2}, \frac{x^2}{(2 -x)^2}\)~,
\ee
and the definition of $Q_{\nu}$,
\be \label{QF}
Q_{\nu}(z) = \frac{\sqrt{\pi}\ \Gamma(\nu +1)}{\Gamma(\nu +3/2) (2 z)^{\nu +1}} F\(\frac{\nu}{2} + 1, \frac{\nu}{2} + \frac{1}{2}, \nu+\frac{3}{2}, z^{-2}\)~.
\ee
The result (\ref{JJint}) reproduces (13-22, 2) of \cite{Watson}. 

Now, we would like to consider the choice of $\mathcal{J}_{\nu}$ in (\ref{Jmath}) with coefficients $a_{\nu} = - b_{\nu}/\tan \nu\pi$ and $b_{\nu} = 1$, which gives, 
\be \label{mJ}
\mathcal{J}_{\nu} = \frac{1}{\sin \nu \pi} J_{-\nu}~.
\ee
Taking this choice of $\mathcal{J}_{\nu}$ in (\ref{mJ}) gives
\begin{multline}
\int_0^{\infty}dt\ e^{- \alpha t} J_{\nu}(\beta t) J_{-\nu}(\gamma t)  \\
=
 \frac{\(\beta \gamma\)^{\nu}}{\Gamma(\nu + \frac{1}{2}) \Gamma(\frac{1}{2}) \Gamma(1-\nu) \alpha}\ \int_0^{\pi} d\phi\ \sin^{2\nu} \phi\ \phantom{}_2 F_1\(\frac{1}{2},1,1-\nu, - \frac{\bar{\omega}^2}{\alpha^2}\)
\end{multline}
This integral can be evaluated to yield a combination of hypergeometric functions of the type $\ \phantom{}_3 F_2$.

\subsection*{Fourier sine and cosine transform of $J_{\nu} J_{\nu}$}
Our starting point is (\ref{JJint}), (see also Eq.~6.612 of \cite{Grad}), 
\be
\int_0^{\infty} d x\ e^{- \alpha x} J_{\nu}(\beta x) J_{\nu}(\gamma x) = \frac{1}{\pi \sqrt{\gamma \beta}} Q_{\nu - 1/2}\Big(\frac{\alpha^2 + \beta^2 + \gamma^2}{2 \beta \gamma}\Big)~,
\ee
where  $\text{Re}(\alpha\pm i \beta \pm i \gamma)>0,\ \gamma>0,\ \text{Re } \nu > - \frac{1}{2}$. 

The Legendre function $Q_{\nu}(z)$ has a branch cut on the real axis running from $-\infty<z<1$. 
We would like to start with $\alpha$ real and continue it to imaginary values. We write $\alpha = |\alpha| e^{i \theta}$, and we will have $\theta$ evolve from $0$ to $\pi/2$. Alternatively, we will also evolve $\theta$ from $0$ to $- \pi/2$. Also, we will assume $\beta, \gamma$ are real.

We define 
\be 
z= \frac{- a^2 + \beta^2 + \gamma^2}{2 \beta \gamma}~,
\ee
and let $a = |\alpha|$. 
We have that 
\be \label{Qsin}
\int_0^{\infty} d x\ \sin(a x) J_{\nu}(\beta x) J_{\nu}(\gamma x) = \frac{ i}{2 \pi \sqrt{\gamma \beta}}  \(Q_{\nu -1/2}(z+ i \epsilon) - Q_{\nu-1/2}(z - i \epsilon) \)~.
\ee

Consider first $0<a<\gamma - \beta$; this corresponds to $z>1$, which is away from the branch cut. As a result, the right hand side of (\ref{Qsin}) vanishes. Next, consider $\gamma - \beta < a < \gamma + \beta$, which corresponds to $-1<z<1$. For this range of $z$, from (8.13) of \cite{Temme}, 
\be
Q_{\nu}(z+ i \epsilon) - Q_{\nu}(z- i \epsilon) = - i \pi P_{\nu}(z)~.
\ee
Finally, for $\gamma + \beta<a$, which corresponds to  $z<-1$, we use the definition of $Q_{\nu}$ in (\ref{QF}).
Since the hypergeometric function $F(a,b,c,x)$ has a branch cut for $x>1$, the only jump in (\ref{QF}) comes from the $z^{- \nu - 1}$ term. Thus, 
\be
Q_{\nu}(z +i \epsilon) -Q_{\nu}(z-i\epsilon) = 2 i \sin \pi \nu\ Q_{\nu}(-z)~,
\ee
for $z<-1$. 
Collecting everything, we get
\be
\int_0^{\infty} d x\  \sin( a x) J_{\nu}(\beta x) J_{\nu}(\gamma x)= \left\{
\begin{array}{lr}
0\ \ \ \ \ \ \ \ \ \ & z>1 \\
 \frac{1}{2  \sqrt{\beta \gamma}} P_{\nu - 1/2}(z), \ \ \ \ & -1<z<1,  \\ 
 -\frac{\cos (\nu \pi)}{\pi \sqrt{\beta \gamma}} Q_{\nu - 1/2} (-z), \ \ \  & z<-1
 \end{array}\right.
\ee
which matches Eq.~6.672 of \cite{Grad}. 
Also, we find that 
\be \label{QcosI}
\int_0^{\infty} d x\  \cos( a x) J_{\nu}(\beta x) J_{\nu}(\gamma x) =  \frac{1}{\pi \sqrt{\gamma \beta}} \left\{
\begin{array}{lr}
Q_{\nu - 1/2}(z)\ \ \ \ \ \ & z>1 \\
 \tilde{Q}_{\nu - 1/2}(z) &  -1<z<1\\
- \sin (\nu \pi)Q_{\nu - 1/2} (-z), & z<-1
\end{array}\right.
\ee
where $\tilde{Q}_{\nu - 1/2}(z) \equiv \frac{1}{2}\(Q_{\nu-1/2}(z+ i \epsilon) + Q_{\nu- 1/2}(z- i \epsilon)\)$
 (Eq.~8.14 of \cite{ Temme}) and is simply  $Q_{\nu -1/2}(z)$ (as defined by Eq.~\ref{eq:QH}).


\end{document}